\documentclass[fleqn,twoside]{article}
\usepackage{espcrc2,isolatin1}

\usepackage{graphicx}
\usepackage[figuresright]{rotating}

\newcommand{\AmS}{{\protect\the\textfont2
  A\kern-.1667em\lower.5ex\hbox{M}\kern-.125emS}}

\def\eq#1{Eq.~(\ref{#1})}

\def\fig#1{Fig.~\ref{#1}}

\def\cB{\mathcal{B}}
\def\cO{\mathcal{O}}
\def\cP{\mathcal{P}}
\def\cR{\mathcal{R}}
\def\ra{\rangle}
\def\la{\langle}

\def\l{\left}
\def\r{\right}

\def\Tr{\,\mathrm{Tr}}
\def\mev{\mathrm{Me\kern-0.1em V}}
\def\gev{\mathrm{Ge\kern-0.1em V}}

\def\NDR{\mathrm{NDR}}
\def\RI{\mathrm{RI}}

\title{ {\vspace{-3.45cm} \normalsize \hfill \parbox{35mm}{\hfill
      CPT-2004/P.066\\ \phantom{x}\hfill BU-HEP 04-16 }
  }\\[22mm]
  Electroweak penguins and SUSY $K^0$-$\bar K^0$ mixing with Neuberger
  quarks\thanks{Work supported in part by US DOE grants
    DE-FG02-91ER40676 and DE-AC02-98CH10866, EU contracts
    HPRN-CT-2000-00145 and HPRN-CT-2002-00311, and grant
    HPMF-CT-2001-01468.  We thank Boston University and NCSA for use
    of their supercomputer facilities.}\thanks{Combined presentations
    by J.~Howard, L.~Lellouch and C.~Rebbi at {\it Lattice 2004},
    FNAL, USA.}  }

\author{F.~Berruto\address[BNL]{Department of Physics, Brookhaven
        National Laboratory, Upton NY 11973, USA\vspace{-0.15cm}},
        N.~Garron\address[CPT]{Centre de Physique Th\'eorique$^1$, Case
        907, CNRS Luminy, F-13288 Marseille Cedex 9, France\vspace{-0.15cm}},
        C.~Hoelbling\addressmark[CPT]\thanks{Present address: 
        Department of Physics, Bergische Universit\"at Wuppertal, 
Gausstr. 20, D-42119, Germany},
        J.~Howard\address[BU]{Department of Physics, Boston
        University, 590 Commonwealth Avenue, Boston MA 02215, USA
\vspace{-0.1cm}},
        L.~Lellouch\addressmark[CPT],
        S.~Necco\addressmark[CPT],
        C.~Rebbi\addressmark[BU],
        and
        N.~Shoresh\addressmark[BU]\thanks{Now at Harvard University.}}
       
\begin{document}

\begin{abstract}
  We present results for $\Delta I{=}3/2$ and $\Delta S{=}2$ matrix
  elements relevant for CP violation in $K\to\pi\pi$ decays and for
  the $K_S{-}K_L$ mass difference in the standard model and beyond.
  They were obtained with Neuberger fermions on quenched gauge
  configurations generated with the Wilson plaquette action at
  $\beta{=}6.0$ on an $18^3\times 64$ lattice.
  %\vspace{1pc}
\vspace{-0.1cm}
\end{abstract}

\maketitle

\footnotetext[1]{UMR 6207 du CNRS et des universit\'es 
d'Aix-Marseille I, II et du Sud Toulon-Var, affili\'ee \`a la FRUMAM.} 

\setcounter{footnote}{1}    

\section{Introduction}
CP violation in $K\to\pi\pi$ decays and the $K_S$-$K_L$ mass
difference provide important constraints on the standard model (SM)
and extensions, once the relevant hadronic matrix elements are
accurately computed. In the following we present results for the
chirally leading $\Delta I{=}3/2$ contributions to direct CP violation,
and for the $\Delta S{=}2$ matrix elements which contribute to
$K^0$-$\bar K^0$ mixing in SM extensions.  These are obtained using
the same Neuberger propagators as for the spectroscopy work presented
in \cite{lat04_spec}, where details of the simulation are given. We
work with degenerate $u$, $d$ and $s$ quarks of bare masses
$am_q=0.03,\,0.04,\,0.06,\,0.08,\,0.10$. The chiral symmetry of
Neuberger fermions guarantees that the mixing on the lattice is
identical to that in the continuum and that our results are fully
$O(a)$-improved.

\section{Electroweak penguins}
At leading chiral order, the $\Delta I{=}3/2$ contribution to
direct CP violation in $K\to\pi\pi$ is determined by the matrix elements
$\langle(\pi\pi)_{I{=}2}|Q_{7,8}|K^0\rangle$, of the electroweak penguin
operators $Q_{7,8}$. Soft pion theorems
can be used to reduce one of the pions, so that~\footnote{The proportionality
constant is convention dependent as is the overall phase of 
$\langle\pi^+|Q_{7,8}^{3/2}|K^+\rangle$.}
\begin{equation}
\langle(\pi\pi)_{I{=}2}|Q_{7,8}|K^0\rangle\propto \frac1{F_\chi}
\langle\pi^+|Q_{7,8}^{3/2}|K^+\rangle
\label{eq:softpion}
\end{equation}
in the chiral limit, where $F_\chi$ is the chiral limit value of the
pion decay constant ($F_\pi=92\,\mev$) and where $Q_{7,8}^{3/2}$ are
the $\Delta I{=}3/2$ components of $Q_{7,8}$, given by
\[
Q_{7,8}^{3/2}=\frac12\left[(\bar sd)_{V{-}A}(\bar uu)_{V{+}A}
+(\bar su)_{V{-}A} (\bar ud)_{V{+}A}\right.
\]
\begin{equation}
\left.\qquad\qquad\qquad-(\bar sd)_{V{-}A}(\bar d d)_{V{+}A}\right]\ ,
\label{eq:o7832def}
\end{equation}
with $Q_7^{3/2}$ color diagonal and $Q_8^{3/2}$ color mixed. Thus, we
calculate $\langle\pi^+|Q_{7,8}^{3/2}|K^+\rangle$ on the lattice using
Neuberger fermions at finite quark mass and extrapolate the
results to the chiral limit.

Because we work in the isospin limit, eye contractions cancel and
power-divergent mixing with lower dimensional operators is absent.
This greatly simplifies the calculation. We begin by constructing the
following ratios of correlation functions ($a=1$):
\[
\cB_{\Gamma_f\Gamma_i}^{7,8}\equiv \frac{N_{7,8}\sum_{\vec{x_i},\vec{x_f}}
\la J_{\Gamma_f}^{\bar du}(x_f)
Q_{7,8}^{3/2}(0) J_{\Gamma_i}^{\bar us}(x_i)\ra}{\sum_{\vec{x_f}}
\la J_{\Gamma_f}^{\bar du}(x_f)
J_{\gamma_5}^{\bar ud}(0)\ra\sum_{\vec{x_i}}\la J_{\gamma_5}^{\bar su}(0)
J_{\Gamma_i}^{\bar us}(x_i)\ra}
\]
\begin{equation}
\qquad\qquad\qquad
\stackrel{T\gg t_i\gg\frac{T}{2}\gg t_f\gg 1}{\longrightarrow}
B_{7,8}^{3/2}\ ,
\label{eq:corrfnrat}
\end{equation}
with $N_7{=}3$ and $N_8{=}1$. $B_{7,8}^{3/2}$ measures
deviation from the VSA of $\la\pi^+|Q_{7,8}^{3/2}|K^+\ra$ in the chiral
limit. In \eq{eq:corrfnrat}, $J_\Gamma^{\bar q'q}=\bar q'\Gamma\hat q$ 
with $\hat q=(1-aD/2\rho)q$. Similarly,
$Q_{7,8}^{3/2}$ are given by \eq{eq:o7832def}, with
quark fields $q$ replaced by $\hat q$ and with the appropriate
rewriting in terms of Euclidean Dirac matrices.

\begin{figure}[t]
\includegraphics*[width=7.3cm]{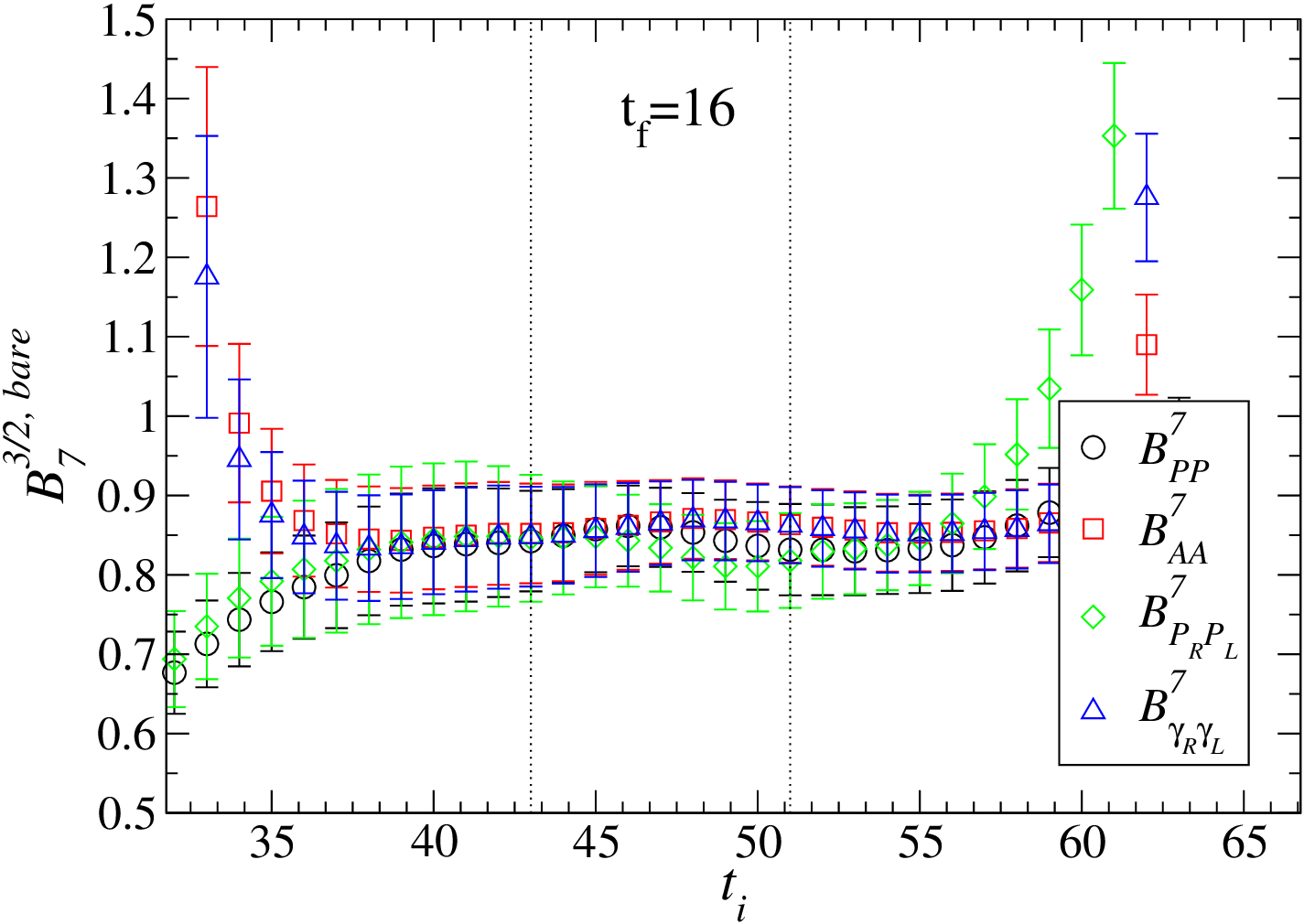}
\vspace{-1.0cm}
\caption{Dependence on initial time $t_i$ of the correlation function
ratio $\cB_{\Gamma_f\Gamma_i}^7$ for a variety of sources and sinks
at fixed final time $t_f=16$ and for $am_q=0.04$.
\vspace{-0.4cm}}
\label{fig:bratvst}
\end{figure}

To quantify possible unwanted contributions from finite-volume zero
modes, we vary the source and sink by letting $\Gamma_{i,f}$ be one of
$\gamma_5$, $\gamma_0\gamma_5$, $(1\pm\gamma_5)$ or
$\gamma_0(1\pm\gamma_5)$. As shown in \fig{fig:bratvst} for our next
to lightest quark mass, $am_q=0.04$, which corresponds roughly to
$m_s/2$, we see no dependence of $\cB_{\Gamma_f\Gamma_i}^7$ on source
and sink in the fit region. $\cB_{\Gamma_f\Gamma_i}^8$ displays very
similar behavior. We interpret this as meaning that zero-mode effects
are negligible around the kaon mass. The fit region, $43\le t_i\le 51$
and $13\le t_i\le 21$, is chosen such that the ratios of
\eq{eq:corrfnrat} are asymptotic and free from wrap-around
contributions. $\cB_{\gamma_5\gamma_5}^{7,8}$ yield the results for
$B_{7,8}^{3/2,bare}$ with the smallest error bars and we take these as
the starting point for subsequent analysis.

\medskip

Before proceeding with the chiral extrapolation of the matrix
elements, we have chosen to renormalize them. We do so
non-perturbatively in the RI/MOM scheme following
\cite{Martinelli:1995ty}.  $Q_7$ and $Q_8$ mix
under renormalization.  We fix gluon configurations to Landau
gauge and numerically compute the relevant amputated forward
quark $q$ Green functions with legs of momentum $p=\sqrt{p^2}$: 
$\Lambda_{Q_{7,8}}(m_q,p^2)$. Then we
determine the renormalization constants by requiring that the
renormalized vertex functions have their tree-level values. Thus,
we define the ratio:
\begin{equation}
\cR^{\RI}_{ij}(m_q,p^2)\equiv Z_A^2\frac{\Tr\l\{\Lambda_V(m_q,p^2)
\cP_{V}\r\}^2}{\Tr\l\{\Lambda_{Q_j}(m_q,p^2) \cP_{Q_i}\r\}}\ ,
\end{equation}
where $i,j\in\{7,8\}$ and the $\cP_\cO$ are normalized projectors onto
the spin-color structure of tree-level $\cO=Q_{7,8},V$. We extrapolate
these ratios to $m_q=0$ and fit the results to the OPE form (including
discretization error terms) \cite{Garron:2003cb}:
\[
\cR^{\RI}_{ij}(0,p^2)=\cdots+\frac{A_{ij}}{p^2}+U^{\RI}_{ik}(p^2)
Z^{RGI}_{kj}
\]
\begin{equation}
\qquad\qquad\qquad+B_{ij}(ap)^2+\cdots\ ,
\label{eq:RRIOPE}
\end{equation}
where $U^{\RI}_{ik}(p^2)$ describes the running of the renormalization
constants in the RI/MOM scheme,
$Z_{ij}^{\RI}(p^2)=Z^{RGI}_{kj}U^{\RI}_{ik}(p^2)$,
implemented at 2-loops \cite{Buras:1990fn,Ciuchini:1995cd} with
$N_f=0$ and $\alpha_s$ from \cite{Capitani:1998mq}. For
$B$-parameters, of course, the VSA's must be appropriately
renormalized.

\begin{figure}[t]
\includegraphics*[width=7.3cm]{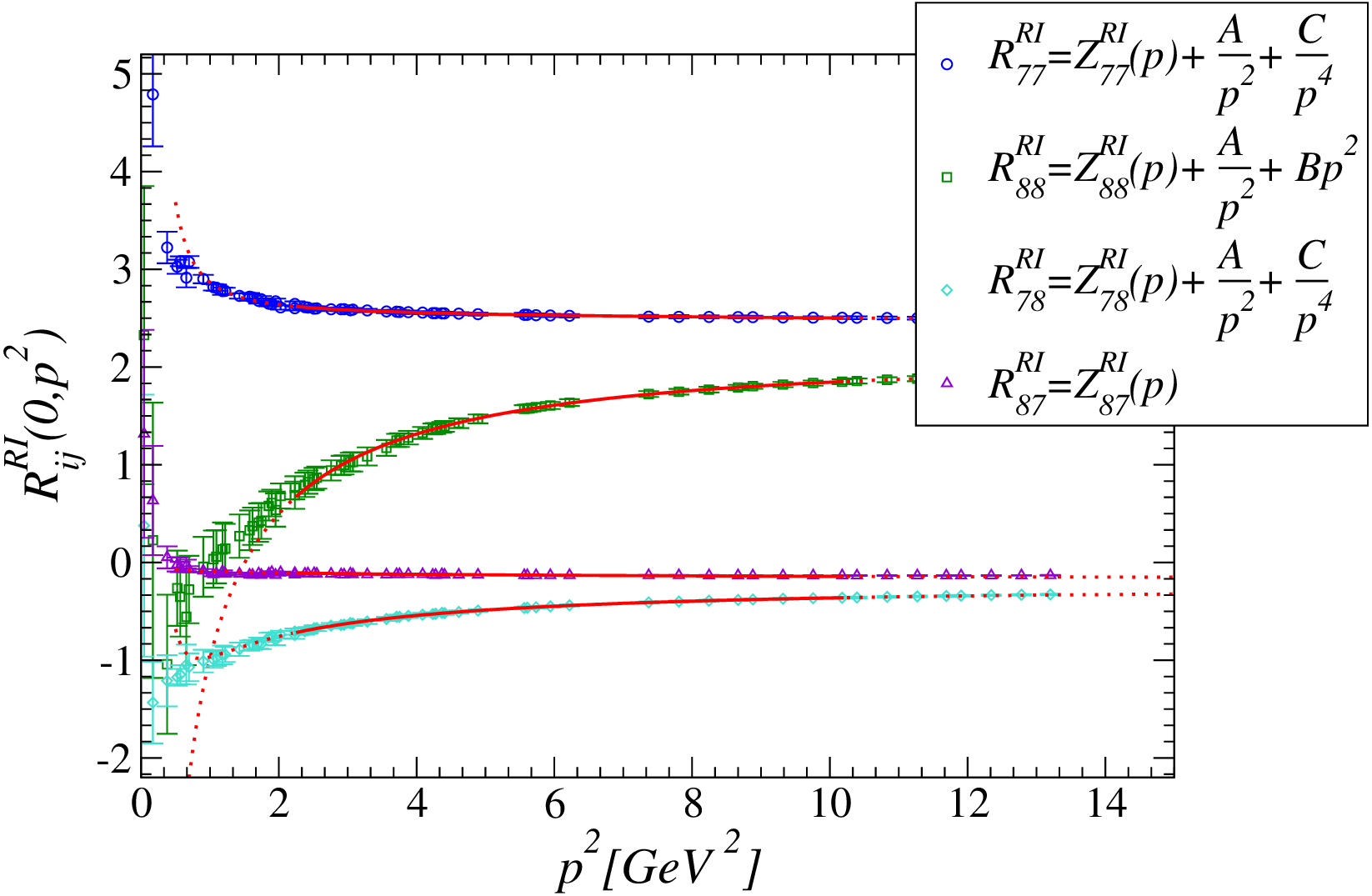}
\vspace{-1.0cm}
\caption{$p^2$-dependence of $\cR^{\RI}_{ij}(0,p^2)$.\vspace{-0.4cm}}
\label{fig:Q78renorm}
\end{figure}

In \fig{fig:Q78renorm} we show the $p^2$-dependence of the ratios
$\cR^{\RI}_{ij}(0,p^2)$ and the fits, for $1.5\,\gev\le p\le
3.2\,\gev$, to the OPE expressions of \eq{eq:RRIOPE} to the orders
displayed in the caption.

\medskip

\begin{figure}[t]
\includegraphics*[width=3.7cm]{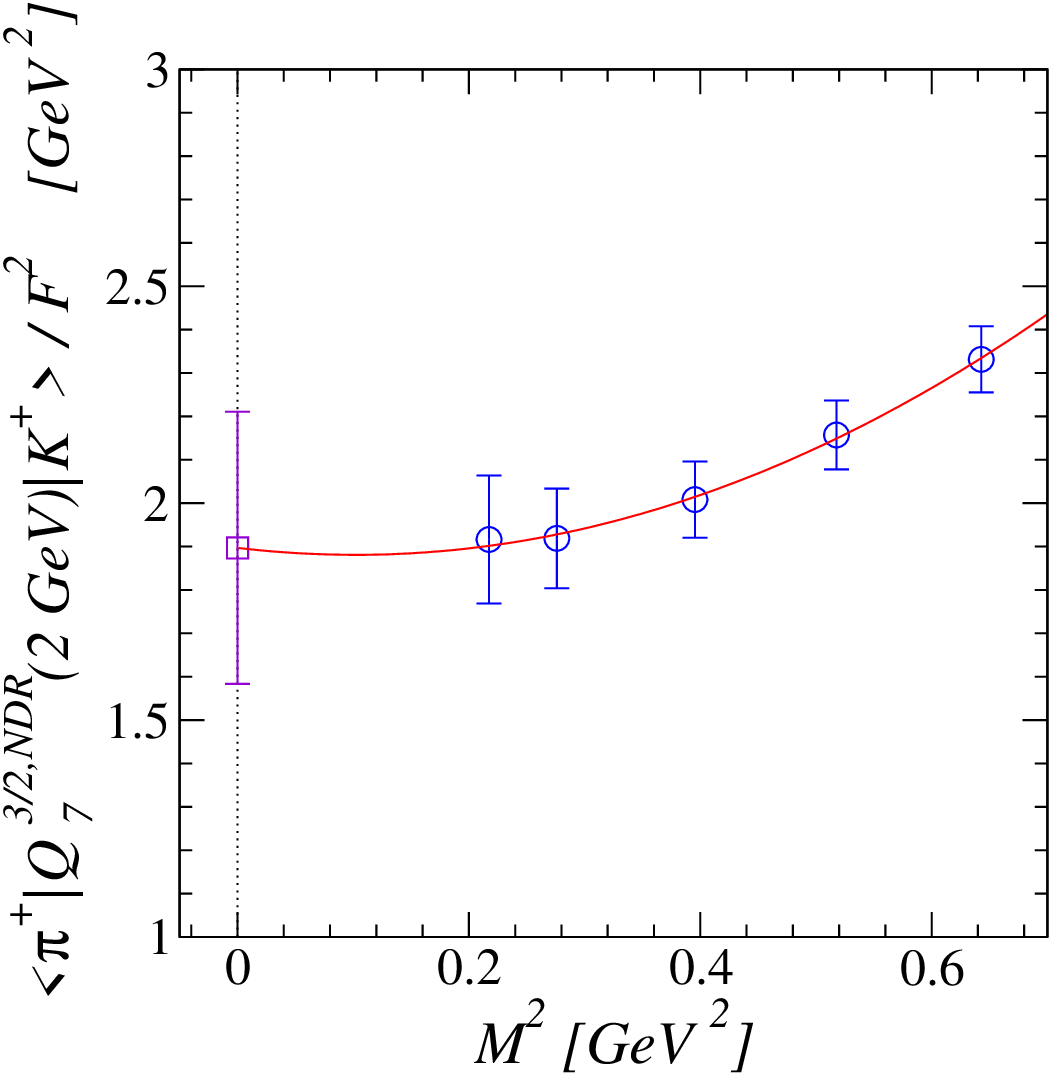}
\includegraphics*[width=3.7cm]{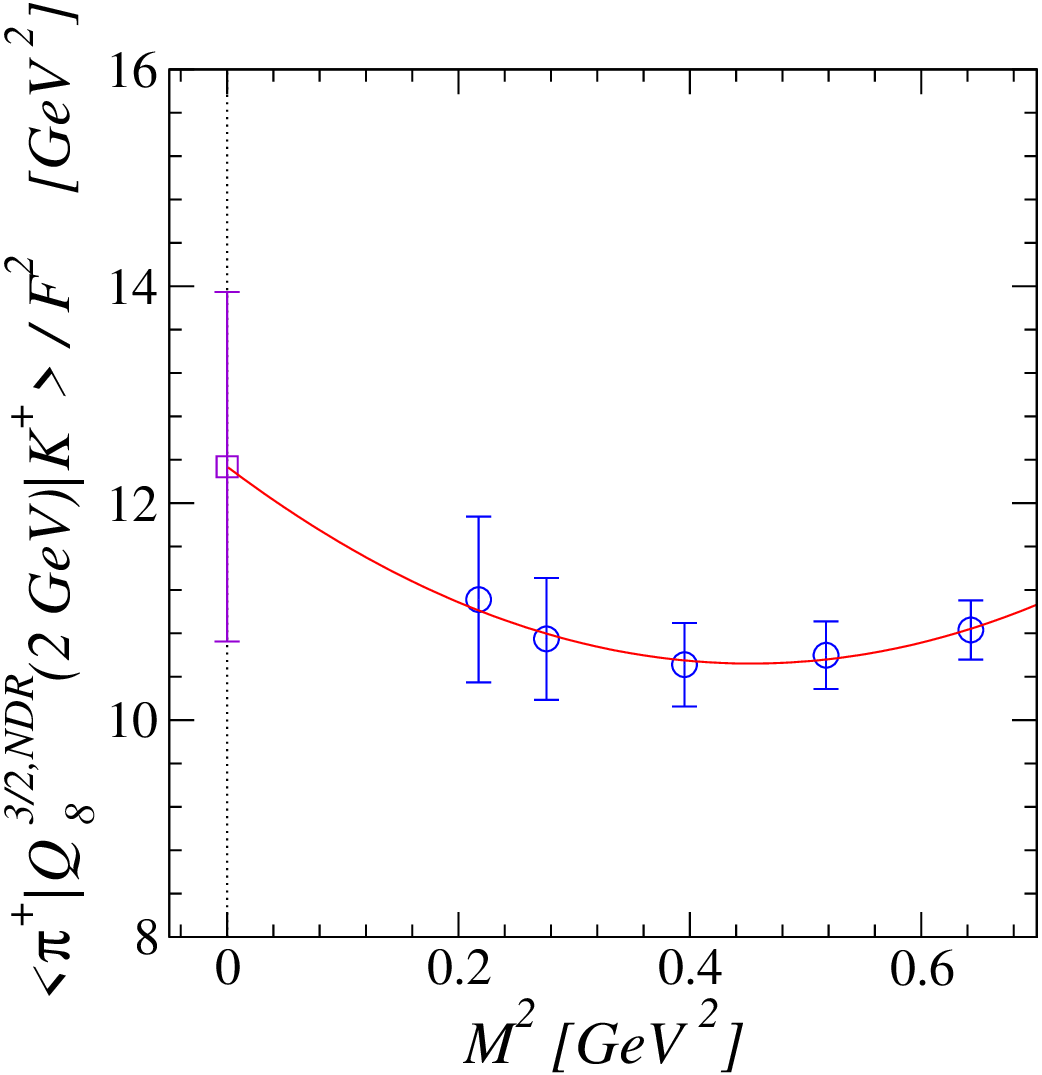}
\vspace{-1.3cm}
\caption{$\la\pi^+|Q_{7,8}^{3/2}|K^+\ra/F^2$ 
in the $\NDR$ scheme at
  $2\,\gev$ are chirally extrapolated.\vspace{-0.4cm}}
\label{fig:D78xextrap}
\end{figure}

We now turn to the chiral extrapolation. Here, complications arise due
to the fact that we are working in the quenched approximation. Indeed,
the VSA's of $\la\pi^+|Q_{7,8}^{3/2}|K^+\ra$ are ill-defined in the
chiral limit due to the presence of quenched chiral logs
\cite{Sharpe:1992ft}. We could consider
$\la\pi^+|Q_{7,8}^{3/2}|K^+\ra/F^4$, with $F$ the decay constant of
our degenerate $K^+$ and $\pi^+$. Instead, we choose to extrapolate
$\la\pi^+|Q_{7,8}^{3/2}|K^+\ra/F^2$, which have
similar mass dependences in the quenched and unquenched theories.
Indeed, at 1-loop in $\chi$PT, 
\begin{equation}
\frac{\la\pi^+|Q_{7,8}^{3/2}|K^+\ra}{F^2}=
\alpha_{7,8}\l[1+\beta_{7,8}
\frac{M^2}{(4\pi F_\chi)^2}\r]\ ,
\label{eq:D78xpt1loop}
\end{equation}
for $N_f{=}3$ \cite{Gasser:1984gg,Golterman:2001qj} {\em and} $N_f{=}0$
\cite{Sharpe:1992ft,Golterman:2001qj}. This similarity in mass dependence
may be an indication that some quenching errors cancel in these
ratios. These ratios have the added
advantage that they are free of chiral logarithms at 1-loop, and
should hence have smoother chiral behaviors. We show the corresponding
chiral extrapolations in \fig{fig:D78xextrap}, where we have added to
\eq{eq:D78xpt1loop} terms proportional to $(M/4\pi F_\chi)^4$ to
account for higher orders. We obtain, in the $\NDR$ scheme at
$2\,\gev$
\begin{eqnarray}
\lim_{m_q\to 0}\frac{\la\pi^+|Q_{7}^{3/2}|K^+\ra}{F^2}=1.9\pm0.3\pm
\mbox{??}\,\gev^2\label{eq:D7res}\\
\lim_{m_q\to 0}\frac{\la\pi^+|Q_{8}^{3/2}|K^+\ra}{F^2}=12.\pm 2.\pm
\mbox{??}\,\gev^2\ ,\label{eq:D8res}
\end{eqnarray}
where $\mbox{??}$ stands for systematic errors which have yet to be
determined. We postpone to a later publication the comparison of our
results with non-lattice
\cite{Narison:2000ys,Bijnens:2001ps,Cirigliano:2002jy,Friot:2004ba}
and quenched Wilson \cite{Donini:1999nn}, domain-wall
\cite{Noaki:2001un,Blum:2001xb} and overlap \cite{DeGrand:2003in}
lattice results.

\section{$\Delta S{=}2$ transitions beyond the SM}

In extensions of the SM, analysis of $K^0$-$\bar K^0$ mixing
generically requires knowledge of $\la\bar K^0|O_i|K^0\ra$
with
\begin{equation}
\begin{array}{rcl}
O_1&=&[\bar s d]_{V-A}[\bar s d]_{V-A}\\
O_{2,3}&=&[\bar s d]_{S-P}[\bar s d]_{S-P}\qquad\mathrm{(unmix,mix)}\\
O_{4,5}&=&[\bar s d]_{S-P}[\bar s d]_{S+P}\qquad\mathrm{(unmix,mix)}
\end{array}
\end{equation}
where: 
\begin{itemize}
\item $\la\bar K^0|O_1|K^0\ra$ is the SM contribution, best given in terms 
of 
$$
B_K=\frac{3}{16}\frac{\la\bar K^0|O_1|K^0\ra}{F_K^2M_K^2}
$$
\item $\la\bar K^0|O_{5,4}|K^0\ra=2\times \la\pi^+|Q_{7,8}^{3/2}|K^+\ra$
for $m_s=m_u=m_d$
\item $\la\bar K^0|O_i|K^0\ra$ is $ O(p^0)$ in the chiral expansion
  for $i\ne 1$.
\end{itemize}
The last point indicates that non-SM matrix elements are expected to
be larger than $\la\bar K^0|O_1|K^0\ra$, which is $O(p^2)$. To make
this explicit, we define the ratios ($i{=}2,\cdots,5$)
\begin{equation}
R_i^\mathrm{BSM}(M^2)
\equiv\l[\frac{F_K^2}{M_K^2}\r]_{expt}\l[\frac{M^2}{F^2}
\frac{\la\bar K^0|O_i|K^0\ra}{\la\bar K^0|O_1|K^0\ra}\r]_{lat},
\end{equation}
where BSM stands for ``beyond the SM'' and where $M$
and $F$ are the mass and ``decay constant'' of the lattice $K^0$.
These ratios have a number of advantages:
\begin{itemize}
\item $\l[\quad\r]_{lat}$ is dimensionless
\item $\l[\quad\r]_{lat}$ is finite in the chiral limit
\item $R_i^\mathrm{BSM}(M_K^2)$ measures directly the ratio of BSM 
to SM contributions.
\end{itemize}

We obtain the required matrix elements from ratios of 3-point to two
2-point functions, as we did for $\la\pi^+|Q_{7,8}^{3/2}|K^+\ra$.
Pseudoscalar sources and sinks were used, except for $\la\bar K^0|O_1|K^0\ra$
where left-handed current sources and sinks were taken to completely
eliminate zero-mode contamination \cite{Garron:2003cb}. We also
perform an RI/MOM non-perturbative renormalization as above: $O_1$
renormalizes multiplicatively; $O_{2,3}$ mix; the $O_{4,5}$ mixing
matrix was already shown.

\begin{figure}[t]
\includegraphics*[width=7.3cm]{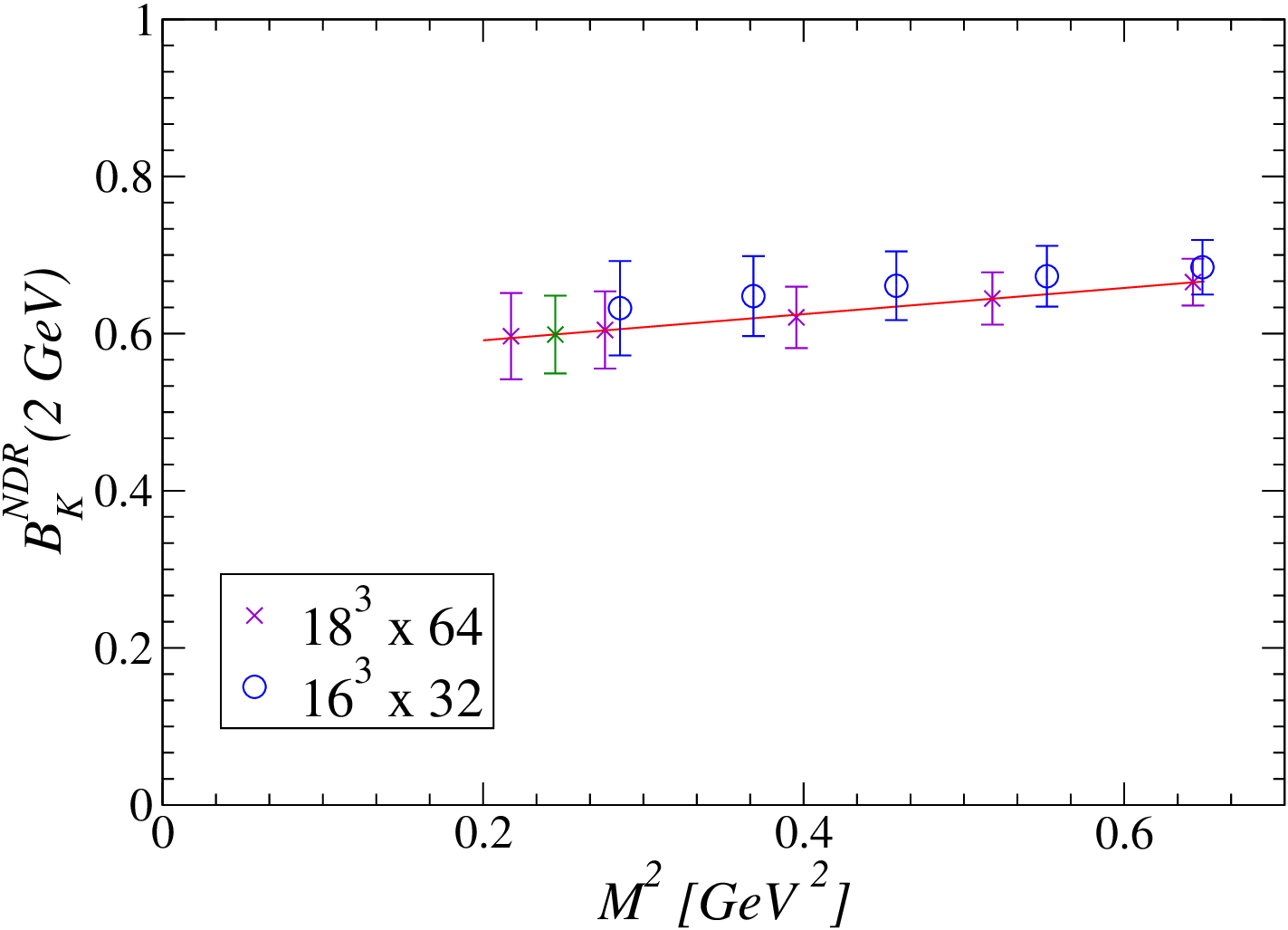}
\vspace{-1.0cm}
\caption{$B_K^{\NDR}(2\,\gev)$ vs $M^2$ at $\beta=6.0$ on $18^3\times 64$
and $16^3\times 32$ \cite{Garron:2003cb} lattices.\vspace{-0.4cm}}
\label{fig:BKvsM2}
\end{figure}

In \fig{fig:BKvsM2}, we present new results for $B_K^{\NDR}(2\,\gev)$
as a function of pseudoscalar mass squared $M^2$, together with our
published results obtained at the same lattice spacing but on a
smaller $16^3\times 32$ lattice \cite{Garron:2003cb}. The results are
entirely compatible, indicating that finite-volume effects are small
around the physical kaon mass.  We find
$B_K^{\NDR}(2\,\gev)=0.60(5)(1)$, which is to be compared with the
value of 0.63(6)(1) that we found on the smaller lattice. Our results
are compatible with the quenched benchmark result of JLQCD
\cite{Aoki:1998nr}, as well as with the overlap results of
\cite{DeGrand:2003in}.

\begin{figure}[t]
\includegraphics*[width=7.3cm]{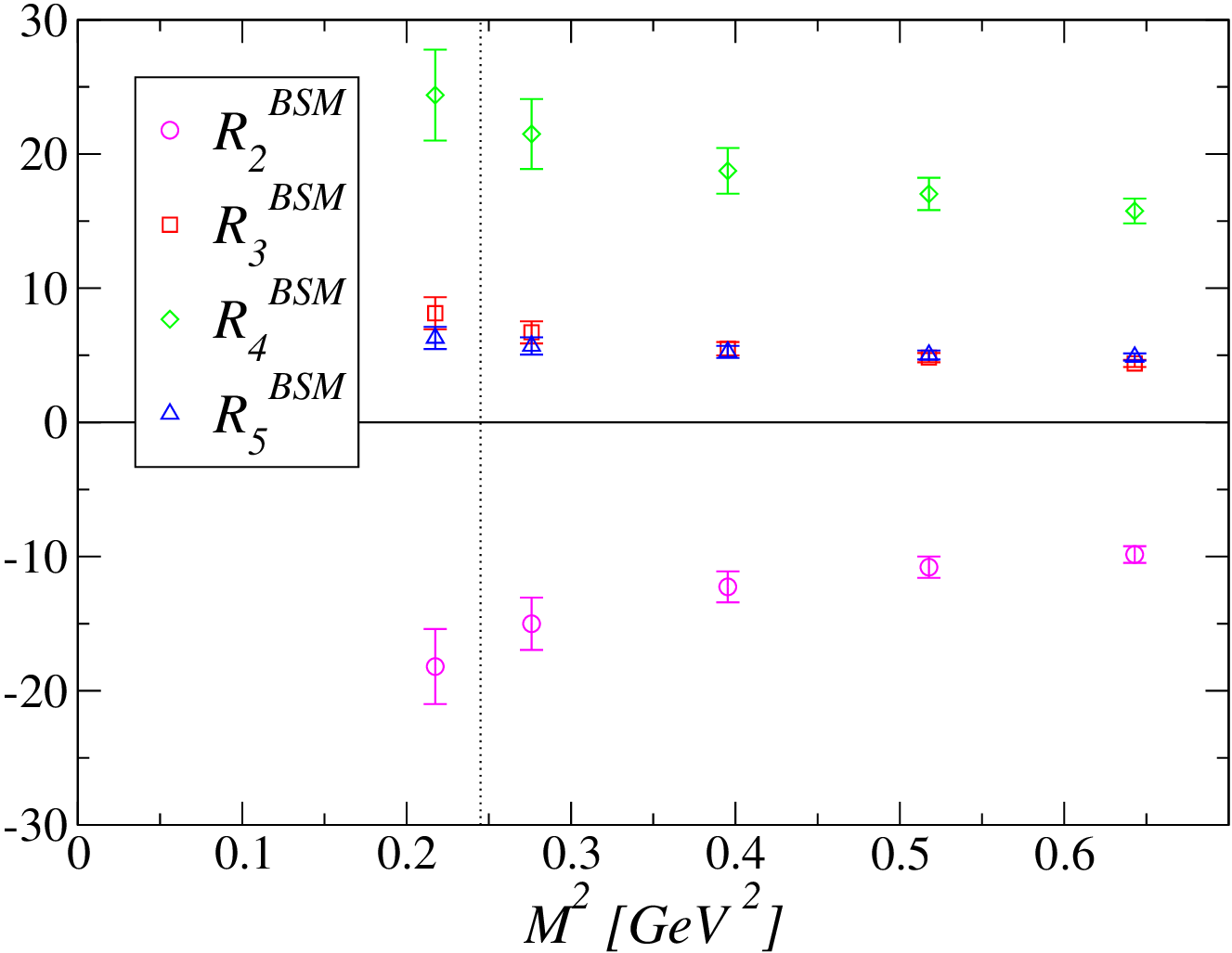}
\vspace{-1.0cm}
\caption{$R_i^\mathrm{BSM}(M^2)$ in the RI/MOM scheme at $2\,\gev$ vs 
$M^2$. The vertical line marks $M^2{=}M_K^2$.\vspace{-0.4cm}}
\label{fig:RIBSMvsM2}
\end{figure}

In \fig{fig:RIBSMvsM2} we plot results for
$R_i^\mathrm{BSM}(M^2)$ in the RI/MOM scheme at $2\,\gev$ as a
function of $M^2$. We find that the BSM matrix elements are enhanced
at $M{=}M_K$ by factors ranging from approximately 5 to 20
compared to the chirally suppressed SM one. This enhancement is
significantly larger than the one observed in the quenched,
Wilson fermion calculation of \cite{Donini:1999nn}.

\section{Outlook}

We are investigating systematic errors on the results
presented above. In particular, 
we are performing the same calculation at $\beta=5.85$ on a
$14^3\times 48$ lattice to quantify discretization errors.

\bibliographystyle{my-elsevier}

\bibliography{wme_lat04}

\begin{thebibliography}{10}
\expandafter\ifx\csname url\endcsname\relax
  \def\url#1{\texttt{#1}}\fi
\expandafter\ifx\csname urlprefix\endcsname\relax\def\urlprefix{URL }\fi
\providecommand{\eprint}[2][]{arXiv:#2}

\bibitem{lat04_spec}
F.~Berruto et~al., presentations by J.~Howard and C.~Rebbi at this conference.

\bibitem{Martinelli:1995ty}
G.~Martinelli et~al., Nucl. Phys. B445 (1995) 81.

\bibitem{Garron:2003cb}
N.~Garron et~al., Phys. Rev. Lett. 92 (2004) 042001.

\bibitem{Buras:1990fn}
A.J. Buras, M.~Jamin and P.H. Weisz, Nucl. Phys. B347 (1990) 491.

\bibitem{Ciuchini:1995cd}
M.~Ciuchini et~al., Z. Phys. C68 (1995) 239.

\bibitem{Capitani:1998mq}
S.~Capitani et~al. (ALPHA), Nucl. Phys. B544 (1999) 669.

\bibitem{Sharpe:1992ft}
S.R. Sharpe, Phys. Rev. D46 (1992) 3146.

\bibitem{Gasser:1984gg}
J.~Gasser and H.~Leutwyler, Nucl. Phys. B250 (1985) 465.

\bibitem{Golterman:2001qj}
M.~Golterman and E.~Pallante, JHEP 10 (2001) 037.

\bibitem{Narison:2000ys}
S.~Narison, Nucl. Phys. B593 (2001) 3.

\bibitem{Bijnens:2001ps}
J.~Bijnens, E.~Gamiz and J.~Prades, JHEP 10 (2001) 009.

\bibitem{Cirigliano:2002jy}
V.~Cirigliano et~al., Phys. Lett. B555 (2003) 71.

\bibitem{Friot:2004ba}
S.~Friot, D.~Greynat and E.~de~Rafael, \eprint{hep-ph/0408281}.

\bibitem{Donini:1999nn}
A.~Donini et~al., Phys. Lett. B470 (1999) 233.

\bibitem{Noaki:2001un}
J.I. Noaki et~al. (CP-PACS), Phys. Rev. D68 (2003) 014501.

\bibitem{Blum:2001xb}
T.~Blum et~al. (RBC), Phys. Rev. D68 (2003) 114506.

\bibitem{DeGrand:2003in}
T.~DeGrand (MILC), Phys. Rev. D69 (2004) 014504.

\bibitem{Aoki:1998nr}
S.~Aoki et~al. (JLQCD), Phys. Rev. Lett. 80 (1998) 5271.

\end{thebibliography}

\end{document}